\documentclass[twocolumn,preprintnumbers,amsmath,amssymb,showpacs]{revtex4}
\usepackage{graphicx}
\usepackage{amsmath, amssymb}
\begin{document}
\title{Buckling of stiff polymer rings in weak spherical confinement}
\author{Katja Ostermeir}
\author{ Karen Alim}
\author{Erwin Frey}
\affiliation{Arnold Sommerfeld Center for Theoretical Physics and Center for NanoScience, Department of Physics,\\ Ludwig-Maximilians-Universit\"at M\"unchen, Theresienstra\ss e 37, D-80333 M\"unchen, Germany}
\date{\today}
\begin{abstract}
Confinement is a versatile and well-established tool to study the properties of polymers either to understand biological processes or to develop new nano-biomaterials. We investigate the conformations of a semiflexible polymer ring in weak spherical confinement imposed by an impenetrable shell. We develop an analytic argument for the dominating polymer trajectory depending on polymer flexibility considering elastic and entropic contributions. Monte Carlo simulations are performed to assess polymer ring conformations in probability densities and by the shape measures asphericity and nature of asphericity. Comparison of the analytic argument with the mean asphericity and the mean nature of asphericity confirm our reasoning to explain polymer ring conformations in the stiff regime, where elastic response prevails.  
\end{abstract}
\maketitle
\section{Introduction}
\begin{figure}[b]
\begin{center}
\includegraphics[width=0.42\textwidth]{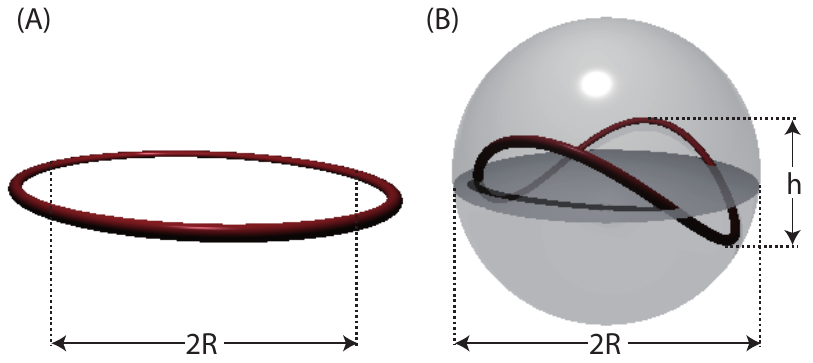}
\caption{(Color online). Dominant shape of a stiff polymer ring without (A) and with (B) spherical confinement. (A) Without confinement the first order bending mode excited by thermal fluctuations induces a planar ellipse that exceeds along its major axis the radius of the corresponding rigid ring. (B) Enclosed by spherical confinement the otherwise planar ellipse is compressed and Euler buckles into a banana-like shape.}
\label{fig_cartoon}
\end{center}
\end{figure}

It is the interplay of elastic energy and entropy that governs the equilibrium form and the dynamics of semiflexible biopolymers. Their competition determines the shape and consequently the function of a biopolymer as a building block in the cytoskeleton \cite{Howard:2001p7507,HBoal:2002p7504} or as an accessible storage medium for genetic information \cite{vandenBroek:2008p281}. Experimental quantification of the elastic and entropic properties of biopolymers often employ confinement, may it be by clamping one end of the polymer \cite{Janson:2004p5619,Pampaloni:2006p5738} or confining the whole polymer into a channel \cite{Reisner:2005p2443,Koster:2008p5732,Choi:2005p7663} or micro-chamber \cite{CosentinoLagomarsino:2007p5633}. In natural conditions the confinement imposed by cell walls and membranes, cell nucleus or viral capsids is approximately spherical. This inspired to use  the rather \emph{weak} confinement of artificial giant vesicles as a versatile and well-controllable model system for the investigation of polymer and polymer bundle characteristics \cite{Elbaum:1996p1593,Limozin:2002p1622,Claessens:2006p142}. Especially but not only in these biomimetic systems, that investigate both biological processes and new nano-biomaterials, polymer rings become of larger and larger importance, stirring theoretical studies of semiflexible polymer rings \cite{SHIMADA:1988p1642,camacho,odijkdna,Panyukov:2001p1656,Rappaport:2006p2076,alimribbon,Norouzi:2008p284,Wada:2009p4066}. DNA on the one hand naturally occurs in ring form \cite{Alberts:2010p6692,Witz:2008p301}  while actin and actin bundles self-assemble into rings under various conditions \cite{Tang:2001p1647,Limozin:2002p1622,Claessens:2006p142,Lau:2009p5614,Sanchez:2010p7457}. Polymer rings are an ideal object to investigate entropic and elastic effects as their topology induces Euler buckling even in weak confinement, where the confining cavity is just equal or a little larger than the average size of the polymer, see Fig.~\ref{fig_cartoon}. Therefore, spherical confinement serves indeed as an excellent tool to investigate the mechanical properties of semiflexible polymer rings and how they are affected due to biological processes under well-defined conditions.   

Within the wormlike chain model semiflexible polymers are characterized by their bending elasticity that opposes the excitation of undulations from thermal fluctuations \cite{rubinstein2003}. Representing a polymer of bending modulus $\kappa$ as a differential space curve $\mathbf{r}(s)$ of length $L$ parametrized by an arc length $s$, its statistical properties are determined by the elastic energy 
\begin{equation}
 \mathcal{H}=\frac{\kappa}{2}\,\int_0^L\,ds \left[\frac{\partial \mathbf{ t}(s)}{\partial s} \right]^2,
 \end{equation}
where $\mathbf{ t}(s) = \partial \mathbf{r}(s)/\partial s$ denotes the tangent vector. The competition of elastic against entropic contributions is reflected in the material specific persistence length, $l_{p}=\kappa/k_b\,T$, which is just the ratio of elastic bending modulus and thermal energy. Comparing this length scale to the total length of the polymer gives a measure of polymer flexibility $L/l_p$. Polymer flexibility easily varies; therefore, our present study takes it as a variable parameter within the region of stiff polymers. It was shown that polymer rings due to their topology effectively behave about five times stiffer than linear polymers, i.e., their stiff regime extends up to $L/l_p\approx 5$  \cite{alimribbon} rendering polymer rings advantageous to study elastic responses.  
 
Semiflexible polymers have previously been the subjects of investigations under conditions of strong confinement where the confining cavity is much smaller than the equilibrium size of the polymer. These conditions arise in viral capsids and bacterial envelopes and provoked both analytical \cite{ODIJK:2004p7570,Metzler:2004p7103,Katzav:2006p1729,Sakaue:2007p70,Morrison:2009p2460} and simulation studies
\cite{Kindt:2001p2009,LaMarque:2004p7291,Ali:2006p7590,Jun:2006p2475,Petrov:2007p7579}. While most studies focus on linear polymers viral DNA may indeed be circular as taken into account for the  investigation of knotting probabilities of polymer rings in strong confinement \cite{Micheletti:2008p7295}. Motivated by nanotechnological advances to study polymers in biomimetic systems semiflexible polymers have furthermore been theoretically investigated in channels \cite{Bicout:2001p7680,Wagner:2007p3427,Yang:2007p7678,Cifra:2009p7230,Thuroff:2010p7601}  on spherical surfaces \cite{ODIJK:1993p1661,Spakowitz:2003p1615,Cerda:2005p7153,Lin:2007p7235} and on two-dimensional planes \cite{Liu:2008p6951,Drube:2009p3755}. Concerning equilibrium properties it is usually the most likely polymer conformation that is relevant for biological processes and nanotechnological applications.

Polymer configuration and form are well-accessible by shape parameters based on the radius of gyration tensor $Q$, given by
\begin{equation}\label{Q}
Q_{ij}=\frac{1}{L}\int_{0}^{L} ds\,
\mathbf{r}_{i}(s) \mathbf{r}_{j}(s)-\frac{1}{L^2}\int_{0}^{L} ds\,
\mathbf{r}_{i}(s)\int_{0}^{L} ds'\mathbf{r}_{ j}(s').
\end{equation} 
The eigenvalues $\lambda_i$ and the direction of the eigenvectors $\Lambda_i$, $i=1,2,3$, of the radius of gyration tensor determine the
spatial extent of a polymer in space. The degree of asymmetry, denoted
asphericity $\Delta$, is proportional to the normalized variance of the
eigenvalues $\lambda_i$ of $Q$ \cite{aronovitz},
\begin{equation}\label{asp}
\Delta=\frac{3}{2}\frac{ 
\sum_{i=1}^{3}\left[\lambda_i-\bar{\lambda}\right]^2
} { (\sum_ { i=1 }
^ { 3 } \lambda_i )^2},
\end{equation}
where $\bar{\lambda}=\sum_{i=1}^{3}\lambda_i/3 $ denotes the mean extent.
While a spherical symmetric object with $\lambda_i=\bar{\lambda}$ is characterized
by the minimal
value of the asphericity $ \Delta=0$, a spherical asymmetric rod-like object is represented
by its maximal value $\Delta=1$.
To measure the degree of prolateness or oblateness of an object, the nature of
asphericity $\Sigma$ is defined by \cite{cannon}:
 \begin{equation}\label{nat}
\Sigma =\frac{4(\lambda_{1}-\bar\lambda)(\lambda_{2}-\bar\lambda)(\lambda_{3}
-\bar\lambda)}{ \left( \frac{2}{3} 
\sum_{i=1}^{3}\left[\lambda_i-\bar{\lambda}\right]^2\right)^{\frac{3}{2}}}.
\end{equation}
The sign of the nature of asphericity is determined by the product of the deviations of the eigenvalues from their mean and is negative for oblate objects and positive for prolate ones. Ranging from 
$\Sigma = -1$ to $\Sigma =1$ the minimal value of the nature of asphericity is attributed to a fully oblate object such as a disk, while the maximal one is assigned to a fully prolate one such as a rigid rod.

We use these shape measures to investigate the form of stiff polymer rings in weak spherical
confinement imposed by an impenetrable shell. Employing both Monte Carlo simulations and analytical calculations we discern elastic and entropic contributions and faithfully describe the dominant polymer conformation depending on polymer flexibility. In Sec.~\ref{sec_buckling} we develop an analytic argument for the trajectory of the dominant polymer conformation considering both entropic and elastic effects. In Sec.~\ref{sec_projections} we assess polymer configurations in spherical confinement over ranges of flexibilities by simulation generated probability densities. Finally, we compare asphericity and nature of asphericity calculated from our analytic argument to their mean values obtained from simulations in Sec.~\ref{sec_shapes}. In the desired stiff regime our analytic argument explains the observed polymer configurations for any weak spherical confinement. We conclude in Sec.~\ref{sec_conclusion}.
\section{Buckling of an elastic ellipse} 
\label{sec_buckling}
To understand the form of polymer rings in spherical confinement it is insightful to have a description of the mean polymer conformation. As the distribution of stiff polymer configurations is indeed sharply centered around the mean, we develop an analytic argument for the space curve of this dominant polymer configuration depending on the strength of the confinement and polymer flexibility.  Based on this dominant space curve (DSC) the governing polymer form can be understood and assessed by calculating its shape parameters. The successful mapping between DSC and simulation results then also ascertains our fruitful insights into the whole polymer configuration. The DSC of a fluctuating stiff polymer ring arises from the interplay of elastic and entropic forces. We analyze their influence subsequently.  To derive the DSC of a stiff semiflexible polymer ring in weak confinement it is instructive to consider first the DSC of an unconfined polymer ring.

A completely rigid polymer ring of contour radius $R_{c}$ is circular. Subjected to thermal fluctuations, it assumes the shape of a planar ellipse \cite{Alim:2007}, the conformation induced by the first bending mode. Increasing flexibility enhances the eccentricity of the ellipse within the stiff regime. While the major axis of the ellipse grows, the minor axis decreases with the square root of the flexibility $\sqrt{L/l_p}$ \cite{camacho,SHIMADA:1988p1642, odijkdna}. As spherical symmetry is broken, this change in shape yields an increase of entropy and, hence, minimizes the free energy. Thus the DSC of the planar stiff polymer ring can be parameterized by 
\begin{eqnarray} \label{eqn_ellipse}
\begin{split}
x(s)=&R_{c}\left(1-\gamma\sqrt{\frac{L}{l_{p}}}\right) \sin\left(\frac{s}{R_{c}}\right),\\ 
y(s)=&R_{c}\left(1+\gamma\sqrt{\frac{ L}{l_{p}}}\right)\cos\left(\frac{s}{R_{c}}\right),\\ 
z(s)=&0,
\end{split}
\end{eqnarray} 
where $s/R_c\in[0,2\pi]$ here represents the polar angle of the trajectory and $\gamma$ denotes a dimensionless parameter that measures the influence of flexibility. The DSC describes a polymer ring that is deformed from an oblate circle to a more and more eccentric ellipse as the flexibility increases. During this growth of eccentricity the total length of the space curve is not conserved, hence, the model does not predict the overall size of a polymer. This caveat does, however, not prevent successful predictions of the shape parameters. As length-invariant measures the asphericity and the nature of asphericity are only affected by the aspect ratio of the axes. In summary, the elliptical form of a free polymer is an entropic effect that can, however, be translated into an elastic response in confinement. 

Confining a stiff polymer ring inside an impenetrable sphere induces a change in its shape. If the major axis of length $S=2 R_{c}(1+\gamma\sqrt{L/l_{p}})$ exceeds the diameter of the sphere $2R$, no planar ellipse can develop inside a sphere. Instead the major axis and therefore the whole polymer ring is compressed by the rigid confining walls of the sphere into a curved banana-like ellipse as shown in Fig.~\ref{fig_cartoon}. This elastic response of the stiff polymer ring to the confinement results in a $z$-component of the DSC. This additional component  generates a banana-like polymer ring, that can be computed by drawing the analogy to the buckling of an elastic rod.

\begin{figure}[t!]
\begin{center}
\includegraphics[width=0.42\textwidth]{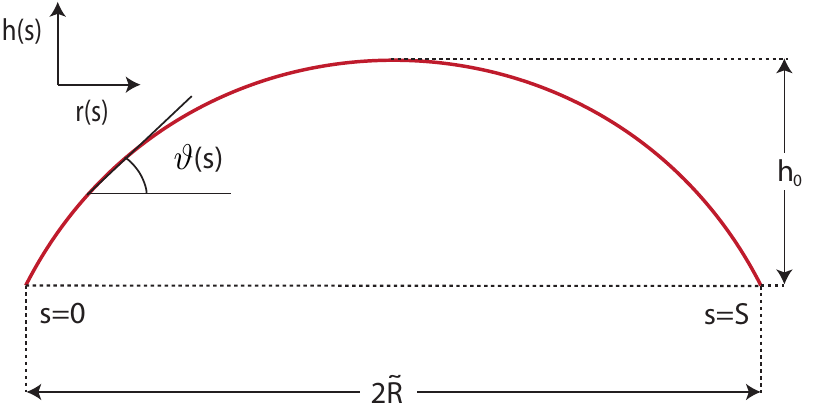}
\end{center}
\caption{(Color online). Buckled rod of length $S$ with maximal height $h_{0}$. The distance between its hinged ends $2\tilde{R}$ determines the rod trajectory parameterized by $\vartheta(s)$, $ s\in [0, S]$ and the maximal height $h_{0}$.}
\label{fig_rod}
\end{figure}
 The conformation of a rod of length $S$ pushing against rigid walls a distance $2\tilde{R}$ apart is equivalent to the shape of a rod of length $S$ being compressed by a constant force $f$ at its hinged ends. As illustrated in Fig.~\ref{fig_rod}, the configuration of a buckled rod of length $S$ can be parameterized by the angle $\vartheta(s)$ between the tangent vector $\mathbf{t}(s)$ and the direction $\hat{\mathbf{r}}$ parallel to the compressing force, where the arc length $s$ runs from $0$ to the length $S$ of the rod. Note, that the rod can rotate and move freely perpendicular to the axis of the force. Reflecting the mirror symmetry of the conformation, the absolute value of the angles at both ends of the rod is equal: $\vartheta(0)\equiv \vartheta_{0}=-\vartheta(S)$. Euler-Lagrange theory predicts the optimal shape of a compressed rod as the state of  minimal elastic energy. The compressive force $f$ adds a potential term to the bending energy of an elastic rod \cite{landauelas},
\begin{equation}
 \mathcal{H}=\int_{0}^{S} ds\left(k_{b}T \frac{l_{p}}{2} \left(\frac{d\vartheta(s)}{ds}\right)^2-
f\cos\left(\vartheta(s)\right)\right),
\label{eqn_elas}
\end{equation}
where we already replaced the bending modulus by its relation to the polymer specific persistence length. As shown in Ref.~\cite{Lee:2007p3733} the minimization of the above elastic energy Eq.~(\ref{eqn_elas}) under the given constraints results in a Euler-Lagrange equation. The minimizing two-dimensional space curve describing the optimal filament shape is then given by its component $r(s)$ along the direction of the force and the component $h(s)$ perpendicular to it, see Fig.~\ref{fig_rod},
\begin{eqnarray}
\begin{split}
r(s)=&-s+\frac{S}{2}+ \frac{S}{\text{K} \left( \sigma \right)}\text{E}\left( \left( \frac{2s-S}{S}\right)
\text{K}\left(
\sigma^2\right), \sigma^2\right),\\
h(s)=&\frac{S\sigma}{ \text{K}\left(\sigma^2\right)}
\left\{1- \text{cn}\left( \left( \frac{2s-S}{S}\right)
\text{K}\left(\sigma^2\right),\sigma^2\right) \right\},
\end{split}
\end{eqnarray} 
where $\sigma$ denotes $\sin(\vartheta_{0}/2)$ and $\text{K}$, $\text{E}$, and $\text{cn}$ are the elliptic integral of the first and second kind and the Jacobi elliptic function, respectively. When the spatial constraint $r(0)-r(S)=2\tilde{R}$ is respected and the elliptic integrals are expanded for small opening angles $\vartheta_{0}$, the maximal  height $h_{0}=|h(S/2)-h(0)|$ depends on the distance between the confining walls $2\tilde{R}$ and the length of the elastic rod $S$ only,
 \begin{equation}
h_{0}(S,\tilde{R})\approx \frac{2}{\pi} S \sqrt{2\left(1-\frac{2\tilde{R}}{S} \right)},
\label{eqn_hmax}
\end{equation}
see Fig.~\ref{fig_rod}. Based on this result the $z$-component of the DSC due to elastic forces can be predicted. Respecting the differential continuity of a buckled ellipse the height modulation function is taken to be a squared sine resulting in the following $z$-component for the DSC in confinement
\begin{equation}
 z(s)=h_{0}(S,\tilde{R})\sin^2\left(\frac{s}{R_{c}}\right).
\end{equation}

In addition to the elastic response due to compression also entropic forces contribute to the DSC of a spherically confined polymer ring. For simplicity, we first assume the confining sphere to be of the same radius $R=R_c$ as the contour radius of the confined polymer ring. In this case any finite temperature causes the major axis of the ensuing ellipse to exceed with its apices the spherical confinement and, hence, forces the polymer to buckle. The elastic bending energy would be smallest if the ellipse's apices both rest on an equatorial plane. Namely, such a configuration maximizes the distance between the apices and, hence, minimizes the curvature of the state. Disregarding rotational symmetry, there is only a single equatorial plane. However,  entropy increases if the apices may rest on any plane instead of just a single equatorial plane. This increase in entropy clearly goes at the expense of stronger bending. Therefore, the magnitude in deviation from the equatorial plane should be related to polymer flexibility. As a good estimate we take the DSC of a polymer ring to nestle half its total height $h_{0}(S,\tilde{R})$ below an equatorial plane and the other half above. Employing Pythagoras law the total length of the major axis is then confined to $\sqrt{R^2-(h_{0}/2)^2}$. Hence, the $y$-component of the DSC follows as 
\begin{equation}
y(s)=\sqrt{R^2-(h_{0}/2)^2}\cos\left(\frac{s}{R_c}\right).
\label{rz}
\end{equation} 
Surely, this nestling below the equatorial plane has also an effect on the buckling height itself, but it is only of second order and, therefore, neglected in the following.

The above equations for the polymer ring's DSC already allow a successful prediction of the mean shape of a polymer ring when the confining radius equals the contour radius $R=R_c$. Next our analysis is extended to larger radii to enable a full description for any kind of confining radius between $R=R_c$ and $R=\infty$. At small flexibilities the DSC of a polymer ring is not supposed to be affected by spherical confinement. As the major axis of the ensuing ellipse does not yet stretch beyond the confining walls, a planar ellipse forms as described by Eq.~(\ref{eqn_ellipse}). Intuitively one would guess that confinement effects become noticeable once the length of the DSC's major ellipse equals the radius of the sphere $R_c(1+\gamma\sqrt{L/l_p})=R$. However, the broadness of the distribution of states makes confinement affect the DSC even before the major vertices of the DSC's planar ellipse encounter the sphere's shell. So far our considerations only described the DSC as the mean space curve irrespective of the broadness of the distribution of states. However, the fraction of configurations with longer major axis forces the DSC to buckle at lower flexibilities than expected. This results in an effectively reduced radius of the sphere, which we account for by choosing $\tilde{R}=R-(1-\alpha)(R-R_{c})$.  For the lower limiting case $\alpha=0$ the effective confinement $\tilde{R}=R_c$ instantaneously affects the DSC irrespective of the true radius $R$, while for the upper limiting case $\alpha=1$ only the encounter of the DSC major axis with the real confinement $\tilde{R}=R$ causes an elastic response. Hence, $\alpha$ denotes the percentage of how much below the real confinement radius $R$ statistically confinement affects the DSC. $\alpha$ is like $\gamma$ a numerical parameter to be determined from simulation data.

Together these two entropic effects and the elastic buckling determine the DSC of a polymer ring of contour radius $R_c$ in spherical confinement of radius $R\geq R_c$. For small flexibilities a planar ellipse develops described by Eq.~(\ref{eqn_ellipse}), that is unaffected by the confinement. This regime extends up to  $R_c(1+\gamma\sqrt{L/l_p})\leq R-(1-\alpha)(R-R_c)$. For larger flexibilities this inequality topples over $R_c(1+\gamma\sqrt{L/l_p})> R-(1-\alpha)(R-R_c)$ and the DSC is described by
\begin{eqnarray}\label{eqn_r}
x(s)=&R_{c}\left(1-\gamma\sqrt{\frac{L}{l_{p}}}\right)\sin\left(\frac{s}{R_{c}}\right),\nonumber\\
y(s)=& R_{c}\sqrt{ \frac{R^2}{R^2_{c}}-\frac{8}{\pi^2} \left(1+\gamma\sqrt{\frac{L}{l_{p}}}\right)\left(\alpha\left(1-\frac{R}{R_{c}}\right)+\gamma\sqrt{\frac{L}{l_{p}}}   \right) }\nonumber \\
&\times\cos\left(\frac{s}{R_{c}}\right),\nonumber\\
z(s)= &\frac{4\sqrt{2}}{\pi} 
R_{c}\sqrt{\left(1+\gamma\sqrt{\frac{L}{l_{p}}}\right)
\left(\alpha\left(1-\frac{R}{R_{c}}\right)+\gamma\sqrt{\frac{L}{l_{p}}}
\right)}\nonumber\\ &\times \sin^2\left(\frac{s}{R_{c}}\right).
\end{eqnarray}
Based on this analytic argument for the DSC of a stiff polymer ring the corresponding shape parameters can be calculated and compared to results from Monte Carlo simulations. Qualitative accordance with our assumptions for the elastic and entropic forces is gained from projections of polymer configurations into two-dimensional planes.
\section{2D Projections of Polymer configurations}
\label{sec_projections}
\begin{figure*}[t]
\begin{center}
\includegraphics[width=.85\textwidth]{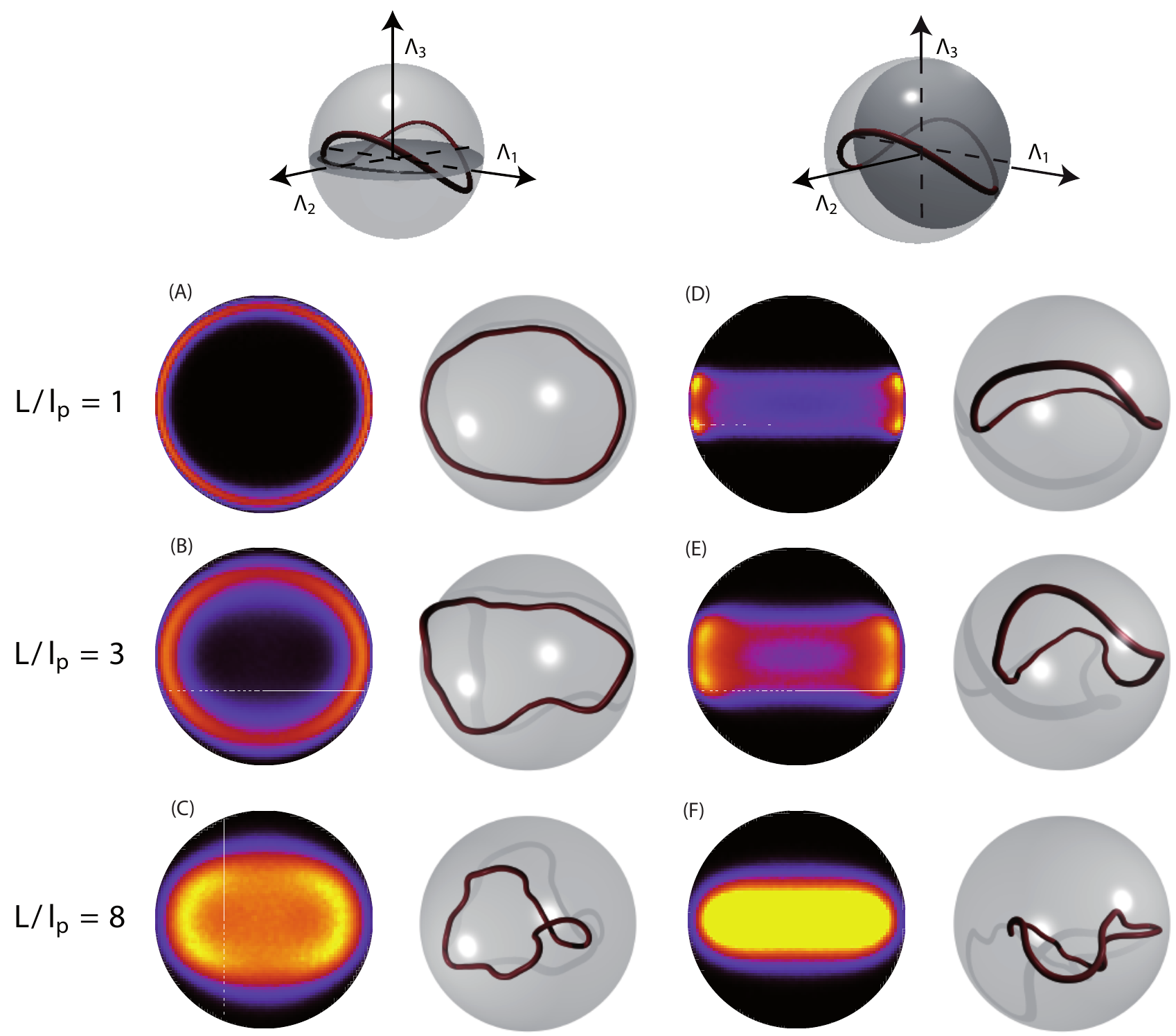}
\caption{(Color online). Probability density and representative snapshots of polymer rings from Monte Carlo simulation data for the flexibilities $L/l_{p}=1, 3$ and $8$. As indicated by the cartoons on top, the polymer configurations in the first column (A-C) are projected onto the plane spanned by the intermediate $\Lambda_2$ and the largest principal axis $\Lambda_1$. In the second column polymer configurations (D-F) are projected onto the plane spanned by the smallest $\Lambda_3$ and the largest principal axis $\Lambda_1$. The gradient in the density of states from high to low is color coded from bright (yellow) to dark (black), the absolute scale of the probability density halves starting from 0.0016 onward as the flexibility increases. The snapshots are chosen such that their asphericity matches the mean configuration of the observed ensemble.}
\label{fig_density}
\end{center}
\end{figure*}
A discretization of the space curve of the polymer ring enables Monte Carlo simulations, which open insights into the governing conformations of polymer rings at different flexibilities.

To simulate a polymer ring of circumference $L$ the Metropolis Monte Carlo method has been employed. The polymer ring is modeled as a discrete polygon, that consists of N segments of fixed length $l$ pointing in the direction $\mathbf{t}$. The elastic energy of a single conformation depends on
the direction between successive segments: $ \mathcal{H}=Nk_{b}T(l_p/L)\sum_{i=1}^{N}(1-\mathbf{t}_{i}\cdot\mathbf{t}_{i+1})$, where the closure of the ring is implemented by periodic boundary conditions $\mathbf{t}_1=\mathbf{t}_{N+1}$. The polymer ring moves through phase space by performing crankshaft moves, restricted by the spherical confinement: Only configurations, which are located entirely inside the rigid walls of the sphere, are considered for averaging. To collect uncorrelated data, only every $10^5$th of those configurations is considered. We sample $10^5$ configurations for each averaged data point, such that the statistical error lies within the ranges of the symbols depicted in our graphs.

To illustrate the form of the polymer rings of radius $R_{c}$ in spheres with $R=R_c$ at different flexibilities, the probability density of polymer configurations are shown in Fig.~\ref{fig_density}. The position vectors of all samples of polymer configurations are mapped on two-dimensional planes spanned by two principal axes of the radius of gyration tensor in Eq.~(\ref{Q}), respectively. Ordering the principal axes $\Lambda_i$, $i=1,2,3$, by the magnitude of their corresponding eigenvalues the largest axis $\Lambda_1$ is taken as reference axis, and the planes spanned together with the intermediate $\Lambda_2$ and the smallest axis $\Lambda_3$ are considered, respectively, to gain insight into the three-dimensional configuration space. Considering a planar ellipse that buckles due to confinement as discussed in Sec.~\ref{sec_buckling}, the plane spanned by the two largest eigenvalues represents the planar ellipse and the plane spanned by the smallest and the largest principal axis corresponds to the height of the buckling polymer relative to the major axis. 

The probability density of polymer configurations in the plane spanned by the largest and the intermediate principal axis in Figs.~\ref{fig_density}(A) - \ref{fig_density}(C) reveals the elliptical character of the mean shape of the polymer ring. At small flexibilities, Figs.~\ref{fig_density}(A) and \ref{fig_density}(B), the polymer trajectories are confined to a narrow rim close to the spherical shell that broadens with increasing flexibility. With growing undulations along the polymer their intermediate axis shortens stronger than the larger one. Hence, polymer configurations resembling an ellipse with higher eccentricity become more probable. Beyond the stiff regime at large flexibilities, Fig.~\ref{fig_density}(C), the polymer ring exhibits compact configurations and looses the character of a planar ellipse. In this semiflexible region, the polymer configurations take a figure-eight shape as indicated by the two yellow semicircles in Fig.~\ref{fig_density}(C). Due to entropic reasons the eight consists of two circles with different sizes for each single polymer configuration \cite{Metzler:2007p325}, therefore, the density distribution is smoothed out in the overlap region of the figure-eight. In the flexible regime, the principal axes shrink further with growing flexibility (data not shown). However, their ratio remains asymmetric to maximize entropy \cite{kuhn1934}.

Studying the density distribution in the plane spanned by the largest and the smallest principal axis we observe bone-shaped probability densities of polymer configurations, see Figs.~\ref{fig_density}(D) and \ref{fig_density}(E). In these projections the density peaks close to the sphere's rim indicate the position, where the elliptically shaped polymer configurations encounter the sphere's shell relative to the equatorial plane. While the largest and intermediate principal axis of the elliptically shaped polymer rings in the stiff regime map onto the major and minor axis of a buckling ellipse, the smallest principal axis points toward the height of the buckling ellipse. Hence, the width of the probability density along the $\Lambda_3$ axis in Figs.~\ref{fig_density}(D) and \ref{fig_density}(E) indicates the maximal buckling height, which in the stiff regime is growing with increasing flexibility. An entirely rigid circular polymer ring would be located in the equatorial plane. With growing flexibility thermal fluctuations force the ensuing ellipse to arch out of the horizontal equatorial plane, forming a bend. Thereby, the major axis of the elliptical polymer ring is clamped below or above the equatorial plane. The position of the ellipses' apices, which pushes against the confining sphere, is marked by the density peaks in the bone-shaped density distribution. The movement of apices' positions away from the equatorial plane with increasing flexibility is an entropic effect taken into account in our analytic argument in Eq.~(\ref{eqn_r}). Beyond the stiff regime, undulations contract the polymers to a degree, that they are no longer forced to undergo Euler buckling but form more and more crumpled configurations also diminishing the polymers' extent along the smallest principal axis.

The entropic and elastic effects observed in the density distributions are in agreement with the analytic argument presented in Sec.~\ref{sec_buckling}. To substantiate these qualitative observations, the observed shapes of polymer rings at different flexibilities in spherical confinement are quantified by the asphericity and the nature of asphericity.
\section{Shapes in spherical confinement}
\label{sec_shapes}
\begin{figure}[tb]
\begin{center}
\includegraphics[width=0.37\textwidth]{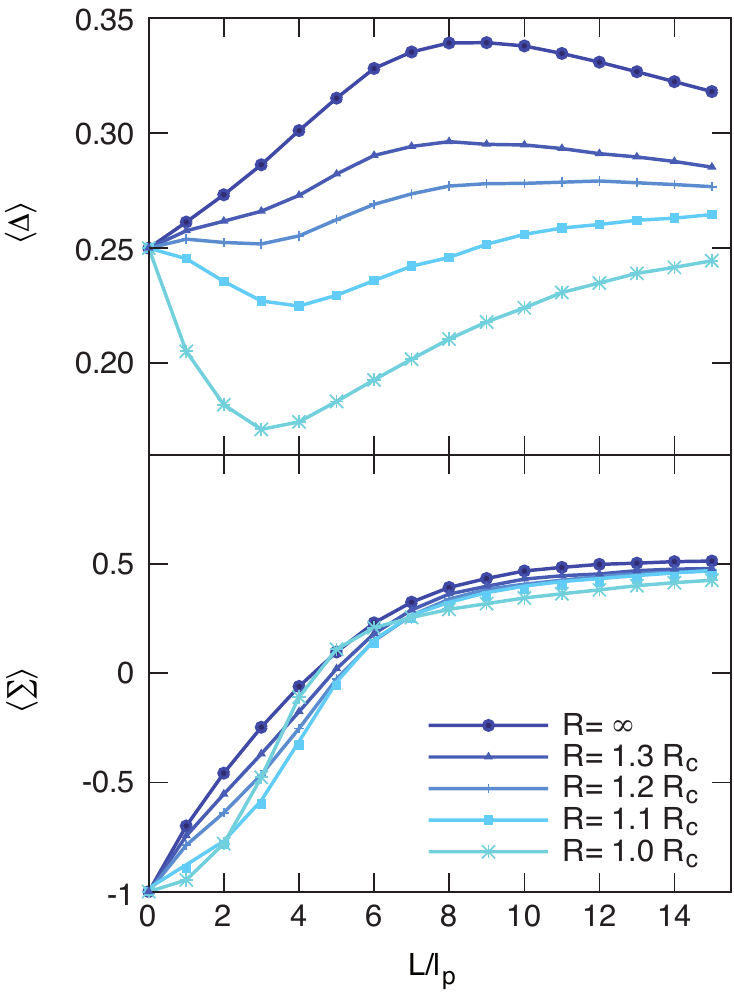}
\caption{(Color online). Monte Carlo simulation data for the mean asphericity $\langle\Delta\rangle$ and the mean nature of asphericity $\langle\Sigma\rangle$ versus flexibility $L/l_p$ for polymer rings of contour radius $R_{c}$, that are confined by impenetrable spheres of radii $R=1.0~R_{c}$ to $R=\infty$. Relatively weak confinement already induces dramatic changes in the shape of polymer rings.}
\label{fig_aspnat}
\end{center}
\end{figure}

\begin{figure*}[tb]
\begin{center}
\includegraphics[width=.87\textwidth]{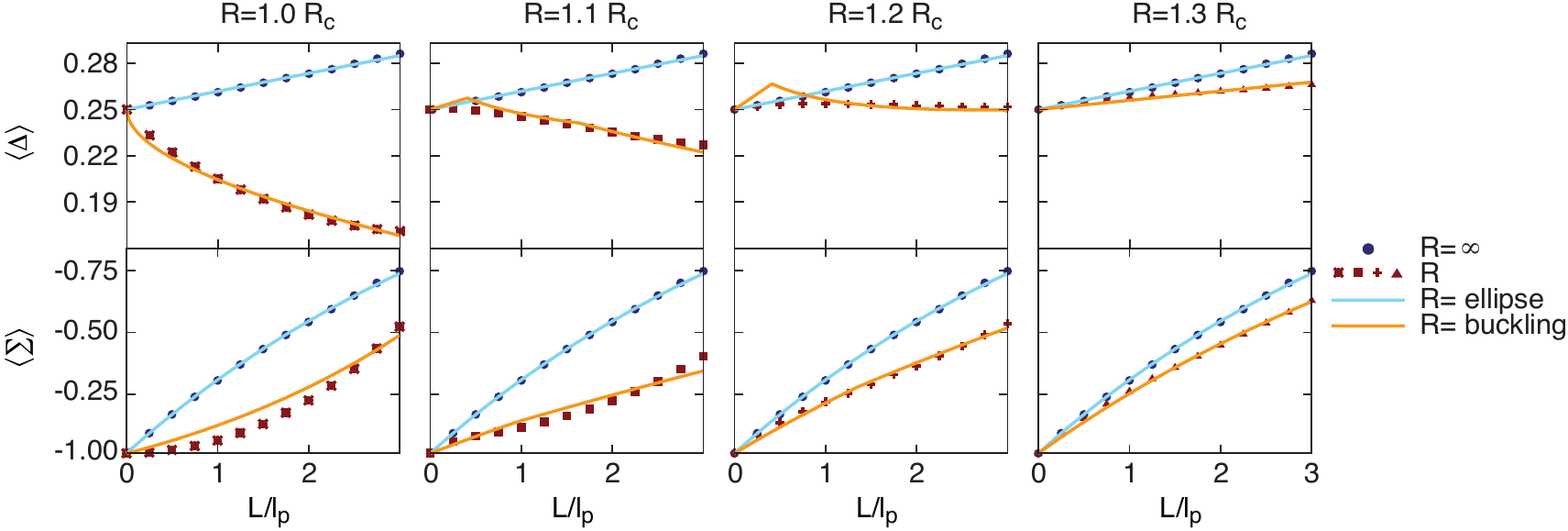}
\caption{(Color online). Comparison of  the mean asphericity $\langle\Delta\rangle$ and mean nature of asphericity $\langle\Sigma\rangle$ versus flexibility $L/l_p$ calculated from our analytical description in Eq.~(\ref{eqn_ellipse},\ref{eqn_r}) (light colors) and from Monte Carlo simulation data (dark symbols) for polymer rings of contour radius $R_{c}$ inside spheres of radii $R=1.0R_{c}$ to $1.3R_{c}$ (lower red series). For reference data and analytical predictions for a free polymer ring are displayed in each diagram as well (upper blue series).}
\label{fig_scaling}
\end{center}
\end{figure*}

The shape of polymer rings is best captured by the asphericity and the nature of asphericity as measures of the extent of asymmetry and the degree of prolateness and oblateness, respectively. Comparing the mean values of these shape parameters for free and confined polymer rings displays the dramatic changes in polymer shape due to weak confinement, as shown in Fig.~\ref{fig_aspnat}. Based on our analytic description for the dominant space curve (DSC) the shapes of polymer rings are rationalized, and by calculating exact values for shape parameters of the DSC we now show that our analytic argument is in agreement with the corresponding Monte Carlo data in the stiff limit, see Fig.~\ref{fig_scaling}.

Starting from $\Delta=0.25$ and $\Sigma=-1$ for a rigid ring the mean asphericity $\langle\Delta\rangle$  and the mean nature of asphericity $\langle\Sigma\rangle$ of a free polymer ring first grow linearly with flexibility $L/l_p$ in the stiff regime due to the increase of the eccentricity of the ensuing planar ellipse \cite{Alim:2007}, see Fig.~\ref{fig_aspnat}. In the semiflexible regime, the free polymer evolves into three-dimensional configurations and undulations lead to crumpling that decreases the variance in spatial extent. Thereby,  the mean asphericity finally decays to the exact value for an infinitely flexible closed Gaussian chain of $\langle \Delta\rangle= 0.2464$ \cite{diehl}. In the course of this transition the polymer form saturates to a prolate, hence, cigar-like shape. In contrast, spherical confinement that is small enough to clamp the largest axis of a polymer ring provokes the mean asphericity to decline in the stiff regime. Only beyond the stiff regime the mean asphericity is observed to grow with increasing flexibility slowly approaching the value of a free polymer ring. Also the linear increase of the mean nature of asphericity $\langle\Sigma\rangle$ of a free polymer in the stiff limit is modified by the confinement and results in a sigmoidal curve progression toward the plateau in the flexible regime.

The decrease in the mean asphericity for  confined polymer rings sets in as the ensuing planar ellipse is restricted by the confining shell and buckles into the third dimension. As the major axis of the polymer increases with flexibility, the buckled polymer conformation gains height and, therefore, looses asphericity. This process progresses up to flexibilities of $L/l_{p}\approx 3$. This marks the end of the stiff regime defined by an elastic buckling. The nature of asphericity displays that in the stiff regime the cigar-like character of the free polymer rings is suppressed by the confinement in favor of more oblate conformations. The inflection point of the mean nature of asphericity reflects the minimum of the asphericity. Increasing the size of the spherical confinement from $R=1.0R_c$ to $R=1.3R_c$ reduces the absolute change in asphericity compared to the free polymer case.  With weaker confinement the onset of the decline of the asphericity is shifted to larger values of flexibility, as the ensuing planar ellipse encounters the shell only at higher flexibilities. As the distribution of polymer extents is broadening with increasing flexibilities this transition is smoothed out more if the buckling sets in at higher flexibilities. Also the character of the nature of asphericity changes at the transition, as clearly shown by the Monte Carlo data in Fig.~\ref{fig_scaling}. If the extension of the polymer rings is smaller than the diameter, the nature of asphericity grows linearly. Its sigmoidal character commences at the transition to buckling.

Beyond the stiff regime, $L/l_{p}> 3$, undulations start contracting the buckled ellipse inducing crumpling to increasingly compact configurations. Thereby, the polymer configurations become less affected by their confinement and both shape measures increase toward the value of unconfined polymer rings. However, over the range of flexibilities observed, even the values in the flexible regime remain distinct. Although the majority of polymer conformations is coiled up within the sphere, very elongated configurations are still discarded and the mean values differ from the unconfined case.

Apart from these qualitative considerations on polymer shape our analytical predictions for the dominant space curve (DSC) in Sec.~\ref{sec_buckling} can be quantitatively assessed by comparison to the shape parameters asphericity and nature of asphericity. Our predictions for the shape parameters depend on two parameters, $\gamma$ and $\alpha$,  as given by Eqs.~(\ref{eqn_ellipse}, \ref{eqn_r}); in the case of $R=R_c$ only a single parameter $\gamma$ is needed, as $\alpha=0$ by definition. The results shown in Fig.~\ref{fig_scaling} are obtained by fitting the parameters $\gamma$ \footnote{asphericity: $R=1.0~R_c$, $\gamma=0.070$; $R=1.1~R_c$, $\gamma=0.077$; $R=1.2~R_c$, $\gamma=0.117$; $R=1.3~R_c$, $\gamma=0.045$; nature of asphericity: $R=1.0~R_c$, $\gamma=0.134$; $R=1.1~R_c$, $\gamma=0.052$; $R=1.2~R_c$, $\gamma=0.065$; $R=1.3~R_c$, $\gamma=0.071$;} and $\alpha$ \footnote{$R=1.0~R_c$, $\alpha=0.0$; $R=1.1~R_c$, $\alpha=0.5$; $R=1.2~R_c$, $\alpha=0.375$; $R=1.3~R_c$, $\alpha=0.5$;} to both observables for all degrees of confinement. Different values for $\gamma$ are obtained for the asphericity and the nature of asphericity.  Such as the mean asphericity represents the average shape of all polymer configurations such does the fitted parameter $\gamma$ only reflect an average of a whole distribution of parameters. Now both asphericity and nature of asphericity have differently shaped, broad and highly skewed distributions. Therefore, the different results for the fitted $\gamma$ reflect only a range of possible values. However, the fit to the stronger peaked asphericity may resemble the average growth with flexibility sufficiently well. Altogether, the fitted curve for the DSC of polymer rings in spheres with radius $R=1.0~R_{c}$ and $R=1.3~R_{c}$ is in good agreement with the simulation results. As our analytical argument does not capture the distribution of states, the smooth transition from planar to buckled ellipses for $R \gg R_{c}$ shows deviations. There, our argument exaggerates the transition in a kink for $R=1.1 R_{c}$ and $R=1.2R_{c}$.  Confirming the quality of our DSC prediction the analytic argument for the nature of asphericity even forecasts its sigmoidal character. In the range between $R=1.1R_{c}$ and $R=1.3R_{c}$ the transition is again not fully captured due to the broad distribution of states, however, the dominant character of the nature of asphericity is well reflected. 
Based on our fitted parameters the magnitude of all three principal axes of the DSC can be calculated for all flexibilities up to $L/l_p=3$. Having two different parameters sets at hand only an estimate of the magnitude is accessible. Recalling that we attribute the fitting results for the stronger peaked asphericity a better representation of all possible polymer states, we employ this value of $\gamma$ to predict for example for $R=R_c$  a maximal buckling height at $L/l_p=3$ of about $h_0/R_c\approx 0.7$.  If one extends the polymer model to account for further microscopic properties as for instance for torsional stiffness, the maximal buckling height is expected to be smaller since torsional stiffness increases the elastic energy of out-of-plane deformations leaving in-plane bending unaffected. Hence, polymers with noticeable torsional stiffness would form elliptical shapes due to in-plane modes as observed for wormlike chain polymers and thus be compressed by confining walls, but the resulting buckling would be to less extend. 
\section{Conclusion}
\label{sec_conclusion}
In summary, the shape and conformation of stiff polymer rings of contour radius $R_c$ in any weak spherical confinement $R\geq R_c$ imposed by an impenetrable shell has been analyzed for varying flexibility $L/l_p$. We find that confining a polymer ring induces buckling due to the polymers elastic properties for finite flexibilities $L/l_p\lessapprox3$. We discern elastic and entropic contributions to the form of polymer rings by simulation-derived probability densities and an analytic argument for the dominating polymer trajectory. While the elastic response can be summarized to Euler buckling, the entropic contribution that broadens the number of accessible states, induces three main effects in the stiff regime. First, entropy promotes planar ellipses for any non-zero flexibility, which increase in eccentricity with growing flexibility. If eventually the major axis is compressed by the confining cavity the polymer ring buckles as an elastic response. Here entropy again takes action as it shifts the plane in which the ring is compressed from the energetically favorable equatorial plane to smaller radii. At last, due to the broad distribution of polymer configurations the transition to buckling is premature and smooth. These four effects are sufficient to explain the form of polymer rings in weak spherical confinement as shown by comparison of shape parameters calculated from our analytic description and averaged simulation data.  

Our analytic description, hence, gives a faithful representation of stiff polymer ring conformations and the scaling of the principal polymer axes, especially the scaling of the buckling height, with polymer flexibility. As our analytic argument accounts beyond polymer flexibility for different radii of confinement, the dominant polymer conformation is now available for active control by these two experimentally adjustable factors. Employing our results in biomimetic experiments, biological processes, that are strongly dependent on polymer configuration, can be investigated under well defined conditions. Furthermore, the knowledge of full polymer conformations is one of the first steps to build nano-structures based on biopolymers. 
\begin{acknowledgements}
The financial support of the Deutsche Forschungsgemeinschaft  through SFB 863 and of the German Excellence Initiative via the program "Nanosystems Initiative Munich (NIM)" is acknowledged. K.A.~also acknowledges funding by the Studienstiftung des deutschen Volkes.
 \end{acknowledgements}

\end{document}